\documentclass[twocolumn,grl]{agutex}

\usepackage{graphicx}
\usepackage{amsmath}

\usepackage[disable]{todonotes}
\setkeys{Gin}{draft=false}
\authorrunninghead{BARTLETT AND STEVENSON}

\titlerunninghead{STABLIZED PRECAMBRIAN DAY LENGTH}

\begin{document}

%
%

\title{Analysis of a Precambrian resonance-stabilized day length}
%
%

%
%



\authors{Benjamin C. Bartlett \altaffilmark{1} and David J. Stevenson,\altaffilmark{1}}

\altaffiltext{1}{California Institute of Technology, Pasadena, California 91125, USA}

Accepted for publication in Geophysical Research Letters on 10 May 2016.

%
%


\begin{abstract}
During the Precambrian era, Earth's decelerating rotation would have passed a 21-hour period that would have been resonant with the semidiurnal atmospheric thermal tide. Near this point, the atmospheric torque would have been maximized, being comparable in magnitude but opposite in direction to the lunar torque, halting Earth's rotational deceleration, maintaining a constant day length, as detailed by \citet{zahnle}. We develop a computational model to determine necessary conditions for formation and breakage of this resonant effect. Our simulations show the resonance to be resilient to atmospheric thermal noise but suggest a sudden atmospheric temperature increase like the deglaciation period following a possible "snowball Earth" near the end of the Precambrian would break this resonance; the Marinoan and Sturtian glaciations seem the most likely candidates for this event. Our model provides a simulated day length over time that resembles existing paleorotational data, though further data is needed to verify this hypothesis.

\end{abstract}

%
%

%

\begin{article}

%
%


\section{Introduction}

At some point during the Precambrian, the Earth would have decelerated to the point where it had a rotational period of 21 hours, which would have been resonant with the semidiurnal atmospheric tide, with its fundamental period of 10.5 hours.  At this point, the atmospheric tidal torque would have been comparable in magnitude but opposite in sign to the lunar oceanic torque, which could create a stabilizing effect on the day length, preserving the 21 hour day length until the resonance was broken, as first discussed in \citet{zahnle}.

The question then arises as to how the Earth broke out of its resonance-stabilized day length of 21hr to progress to its current day length of 24hr. In general, any sufficiently large sudden increase in temperature will shift the resonant period of the atmosphere by thermal expansion (resulting in a change of atmospheric column height) to a shorter period, as described in Figure 1, and could potentially break resonance, allowing for Earth to decelerate to longer day lengths. (Alternately, resonance could also be broken by increasing the lunar torque to surpass the peak atmospheric torque by the gradual change of the oceanic $Q$-factor, defined for an arbitrary system as $2\pi\cdot\frac{\text{total energy}}{\text{energy dissipated per cycle}}$, though the very low necessary atmospheric $Q$ factor for resonance to form given the current oceanic torque makes this seem a less likely explanation and is not explored here.)

This study develops a model of resonance formation and breakage that approximately outlines the necessary conditions for this constant day length phenomenon to occur for an extended period of time. In our model of atmospheric resonance, there are effectively three outcomes for this resonant phenomenon.

First, the Earth could have entered a stable resonant state which lasted for some extended period of time before being interrupted by a global temperature increase, such as the deglaciation period following a possible "snowball Earth" event. Specifically, the Sturtian or Marinoan glaciations make good candidates for this breakage event. \citep{pierrehumbert, rooney}

Second, the resonant stabilization could have never occurred, as the $Q$-factor of the atmosphere could have been too low for the magnitude of the atmospheric torque to exceed that of the lunar torque, a necessary condition for a constant day length.

Third, the resonance could have been of no interest, as atmospheric and temperature fluctuations could have been too high to allow a stable resonance to form for an extended period of time.

We discuss the plausibility of each of these scenarios in greater detail below and ultimately conclude that the first scenario is the most likely to have occurred.

\section{Analysis of atmospheric resonance}

The details of the atmospheric tide are quite complex, but the essential features can be appreciated with the following toy model of the torque. (For interested readers, a more complete treatment of the atmospheric tidal problem is given in \citet{lindzenandchapman}, most specifically in Section 3.5.C.)

Given a fluid with column density $\rho_0$ and equivalent column height $h_0$ under gravitational acceleration $g$, with Lamb waves of amplitude $h \ll h_0$ and wavelength $\lambda \gg h_0$, wave speed of $\sqrt{g h_0}$, Cartesian spatial coordinates of $x$, a forced heating term $h_f$, and a damping factor $\Gamma=\frac{1}{t_Q}$ (with $t_Q$ defined as the total energy over power loss of the system, such that $Q=\omega_0 t_Q$), we start with the forced wave equation without drag (we will add this in later):

	\begin{equation} \label{eqnofmotion3}
	\frac{\partial^2 h}{\partial t^2} = g h_0 \frac{\partial^2 h}{\partial x^2} + \frac{\partial^2 h_f}{\partial t^2}
	\end{equation}

We are interested in a heating term of the form $F = F_0 \cos(2\omega t + 2 k x)$, with $F_0$ as the average heating per unit area, $\omega$ as the angular frequency, and $k=\frac{2\pi}{2\pi R_\oplus}$ at the equator, with $R_\oplus$ the Earth's equatorial radius.  Thus, for $C_p$ as the specific heat at constant pressure and $T_0$ as mean surface temperature, we have that $\rho_0 C_p T_0 \frac{dh_f}{dt} = F_0 \cos(2\omega t + 2 k x)$, or:

	\begin{equation}
	h_f = \frac{F_0 \sin(2\omega t + 2kx)}{2 \rho_0 C_p T_0 \omega}
	\end{equation}	
Expressing $h=A\sin(2\omega t + 2kx)$ and defining the equivalent height (which is currently 7.852km \citep{zahnle}, with resonant effects occurring at about 10km) as $h_0 \equiv \frac{4 \omega^2 R_\oplus^2}{\beta g}$, where $\beta$ is the relevant eigenvalue to Laplace's tidal equation. Using present values for $h_0$, $\omega$, $g$, and $k$, we obtain $\beta\approx 0.089$, which agrees swith \cite{lamb}, p. 560. Near resonance, $g h_0 = \frac{4 \omega^2}{\beta k^2}\approx \frac{4 \omega_0^2}{\beta k^2}$, so we obtain via Equation \ref{eqnofmotion3} that:

	\begin{equation}
	A = -\frac{\omega F_0}{2\rho_0 C_p T_0 (\omega_0^2-\omega^2)}
	\end{equation}
	
At the present day, $\omega<\omega_0$, making $A$ negative, so the positive peak of $A\sin(2\omega t+2kx)$ is at $2\omega t + 2kx = -\frac{\pi}{2}$.  At noon ($t=0$), this occurs spatially at $x=-\frac{\pi}{4k}=-\frac{\pi R}{4}$, or $-45^\circ$.  This result determines the sign of the torque, as the mass excess closer to the sun exists such that it is being pulled in the prograde rotational direction.  Note that for period of time where the length of day is less that the resonant period of 21hr, that is, for $\omega>\omega_0$, the resultant torque of $A$ will exert a decelerating effect on the earth.  However, at the point of resonance in question, where the lunar torque is cancelled by the atmospheric torque, $\omega<\omega_0$ by a small factor, corresponding to a day length slightly above 21hr.

Addressing drag in our model, if we assume that any excess velocity formed from the tidal acceleration in the atmosphere is quickly dissipated into the Earth through surface interactions with a damping factor $\Gamma$, and that this surface motion is relatively quickly dissipated into the rotational motion of the entire Earth, as given by \citet{hide}, writing the dissipative Lamb wave forces, we have:

	\begin{equation}
	\frac{\partial v}{\partial t} = -g\frac{\partial h}{\partial x} - v\Gamma \qquad \qquad
	h_0 \frac{\partial v}{\partial x} = - \frac{\partial h}{\partial t} + \frac{\partial h_f}{\partial t}
	\end{equation}
from which we obtain: 

	\begin{equation}
	A = \frac{F_0}{\rho_0 C_p T_0} \cdot \frac{(2\omega - i \Gamma) (4(\omega^2-\omega_0^2) + 2i\omega \Gamma)}{16(\omega^2-\omega_0^2)^2 + 4\omega^2 \Gamma^2}.
	\end{equation}
	
In this model, the imaginary component $\Im(A)$ represents amplitude which would create a force with angle of $\frac{\pi}{2}$ with respect to the sun, and thus does not exert any torque on the Earth.  We need only concern ourselves with the real part $\Re(A)$, then. Thus, we have:

	\begin{equation} \label{Aeqn}
	\Re(A) = \frac{F_0}{2 \rho_0 C_p T_0} \cdot \frac{4\omega(\omega^2-\omega_0^2) + \omega\Gamma^2}{4(\omega^2-\omega_0^2)^2+\omega^2 \Gamma^2}.
	\end{equation} 

Since we know the atmospheric torque to be directly proportional to the atmospheric displacement $A$, we can use the fact that the present accelerative atmospheric torque, $~2.5\times 10^{19}$Nm, is approximately $\frac{1}{16}$ that of the present decelerative lunar torque, $~4\times 10^{20}$Nm, as given in \citet{lambeck}, to scale the atmospheric torque along the curve following $\Re(A)$, giving the total atmospheric torque $\tau_{atm}(\omega)$ as a function of the Earth's rotational frequency, as detailed in Figure 1.

\begin{figure*}
\centering
\noindent\includegraphics[width=\textwidth]{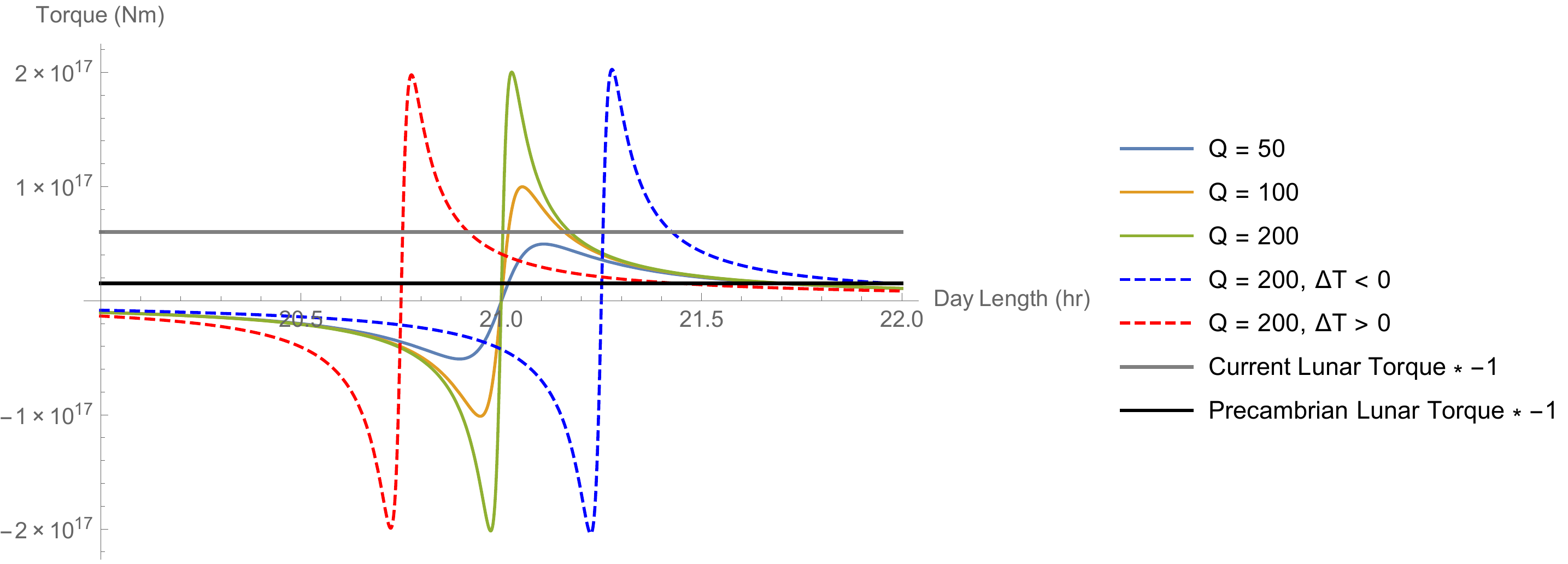}
\caption{Torque values for atmospheric torques assuming various $Q$-factors compared to lunar torque.  Torques are scaled at the 24hr. endpoint such that they have a value $\frac{1}{16}$ that of the lunar torque, while the contour of the curve is determined by the $A$ term derived in section 2. Note that the minimum value of $Q$ required to form a resonance (the value such that its magnitude exceeds the lunar torque) varies linearly with the lunar torque. During the Precambrian, when the lunar torque was thought to be approximately a fourth of its current value \citep{zahnle}, very low values of $Q$ could have permitted resonance to form.}

\label{figure_label}
\end{figure*}

For a sufficiently high atmospheric $Q$, we can see that, starting from a short day length, as the Earth decelerates over time, increasing the length of day, the atmospheric torque increases until it eventually matches the lunar torque, keeping the length of day constant at this stable equilibrium. While there are two day lengths at which the torques are balanced, only the one at lower day length is stable. That is, infinitesimally perturbing the system about the unstable equilibrium will cause the system to move away from equilibrium (to a longer day length).

It should be noted that, to the authors' knowledge, there is little consensus on a value of $Q$ for the atmosphere, though \citet{lindzen} put $Q$ for a period of 10.5 hours at about 30. Regardless, one can reasonably assume it is within the range of $10-500$, so we solve the problems in this paper using all possible values of $Q$ within this range. Ultimately, we establish a critical (relatively low) threshold, dependent on the lunar torque, that $Q$ must exceed for resonance to form - all systems with $Q$ sufficiently past this threshold result in similar results.

\section{Estimation of resonance-breaking conditions}

Before solving the deglaciation timescale problem with a more complete computational model based on the previous section, we detail a less sophisticated analytical solution to approximate the warming timescale necessary to break resonance. We then verify this with our computational model, noting that the key features of the solution are present in both models, albeit at different values.

Given some increase in global temperature $\Delta T$ from an initial "average" temperature $T_0$, we would expect a corresponding increase in atmospheric volume. Since the atmosphere is horizontally constrained, this should result in a nearly linear increase in the column height of the atmosphere. This, in turn, would change the propagation speed of an atmospheric Kelvin wave, given by $v=\sqrt{gh_0}$, and thus the resonance frequency of the atmosphere.  A decrease in global temperature increases the atmospheric resonant frequency (thus decreasing the equilibrium length of day, shifting the curves to the left on Figure 1), while an increase in global temperature decreases this frequency. 

Suppose Earth had progressed to the stable equilibrium point in Figure 1. A large, fast (but non-instantaneous) increase in global temperature could shift this stable equilibrium point to sufficiently lower day lengths such that the unstable equilibrium point would be shifted past Earth's day length, allowing the Earth to freely decelerate to longer day lengths.  This change in temperature would need to be sudden enough that the Earth's rotation could not "track" this change. Additionally, decreasing temperature to its previous value before the Earth has a chance to migrate away from the resonant zone could result in a recapture event.

Let's examine how fast this increase in temperature would need to occur. For a change in resonant frequency to preserve resonance throughout the duration of the change, the rotational frequency of the Earth must track the change in resonance frequency of the atmosphere, so $\frac{d\omega}{dt}\sim\frac{d\omega_0}{dt}$. If we make the simplifying approximation that the torque curve in Figure 1 has zero width (removing any "buffer zone" about the stable equilibrium), then we require $\frac{d\omega}{dt}=\frac{d\omega_0}{dt}$. Since the resonance frequency of the atmosphere is $\omega_0 = \frac{\sqrt{gh_0 / \epsilon_{22}}}{2 R_\oplus}$, and $h_0\propto T$ for temperature $T$, we can express $\omega_0(T)$ as $\omega_0 = \frac{\sqrt{\frac{gh_0}{\epsilon_{22}} \cdot \frac{T}{T_0}}}{2 R_\oplus}$, with $T_0$ the initial temperature.  For any realistic changes in atmospheric temperature, $\frac{T(t)}{T_0} \approx 1$. Denoting the time over which the temperature changes by an amount $\Delta T$ as $t_w$, we obtain:

	\begin{equation}
	\frac{d\omega_0}{dt} =\frac{d T(t)}{dt}\cdot \frac{\sqrt{\frac{gh_0}{\epsilon_{22}} \cdot \frac{T(t)}{T_0}}}{2 R_\oplus T(t)} \approx \frac{\Delta T \sqrt{gh_0/\epsilon_{22}}}{4 t_w T_0 R_\oplus}.
	\end{equation}

Following the amplitude-scaling technique mentioned in the previous section, we know the maximum angular acceleration of the Earth, or the fastest the Earth can "track" changes in $\omega_0$, to be:

	\begin{equation}
	\frac{d\omega}{dt} = \frac{\tau_{\text{atm}}-\tau_{\text{moon}}}{I_\oplus}=\frac{\tau_{\text{moon}}(\frac{A(\omega_{max})}{16\cdot A(\frac{2\pi}{24hr})} - 1)}{I_\oplus}
	\end{equation}
	
where $\omega_{max}$ is the rotational frequency associated with the global maximum of $\tau_{atm}$, and $\frac{2\pi}{24hr}$ is the current rotational frequency of the earth.  Abbreviating $A(\omega_{\text{max}})$ as $A_{\text{max}}$ and $A(\frac{2\pi}{24hr})$ as $A_{24}$, we obtain that:

	\begin{equation}
	\frac{\Delta T \sqrt{gh_0/\epsilon_{22}}}{4 t_w T_0 R_\oplus} = \frac{\tau_{\text{moon}}}{I_\oplus} \left( \frac{A_{\text{max}}}{16 \cdot A_{24}} - 1 \right)
	\end{equation}
	
Since $A_{\text{max}}$ scales very nearly linearly with $Q$, we need only attain one value of $A_{\text{max}}$ and scale it accordingly with $Q$. For example, at $Q=100$, $A_{\text{max}} \approx 27.01\cdot A_{24}$, and at $Q=200$, $A_{\text{max}} \approx 53.78\cdot A_{24}$.  Using the value at $Q=100$, our expression for the minimum $t_w$ over which the temperature can change by $\Delta T$ without breaking resonance becomes:

	\begin{equation}
	t_w \approx \frac{\Delta T I_\oplus \omega_0}{2 T_0 \tau_{\text{moon}}(\frac{27}{16} \frac{Q}{100} - 1)}.
	\end{equation}
	
As shown in Figure 2, this expression indicates asymptotes for stability-preserving $(Q,t_w)$ pairs as $Q\rightarrow 60$ and as $t_w \rightarrow 0$. For a plausible atmospheric $Q$-factor of 100 and a temperature change of $\Delta T = 10K$, any significant change in temperature faster than on the order of $10^8$ years will break resonance.

Note that this model simply provides an upper bound on the time over which the temperature can be changed while preserving resonance and an upper bound on the minimum threshold for $Q$ for resonance to form. This is due to the approximation that the curves in Figure 1 have a half-maximum width of zero and that resonance will be broken Earth's rotation at all deviates from the shifting equilibrium. In reality, the width of the curves provides a buffer for the Earth; for example, very small changes of resonance frequency will not break resonance, as the displacement is not sufficient to put the Earth outside of the stable zone, even if the change occurs instantaneously. Thus, temperature changes will need to occur over a shorter $t_w$ than the values shown in Figure 2 to actually break resonance, and resonance may form with lower values of $Q$. Additionally, the current ratio of atmospheric to oceanic lunar torques is hard-wired in this estimate, overstating the oceanic torque and understating the strength of the resonance. These problems are more precisely addressed with our computational model outlined in the next section. However, this simple model gives us a good idea of the behavior we should expect: an asymptote at low $Q$, $t_w$ scaling approximately with $\frac{1}{k Q - 1}$, and, for sufficiently large $\Delta T$ such that the buffer zone is small in comparison to the induced change in resonant frequency, $t_w$ scaling linearly with $\Delta T$.

\begin{figure}
\centering
\noindent\includegraphics[width=.9\columnwidth]{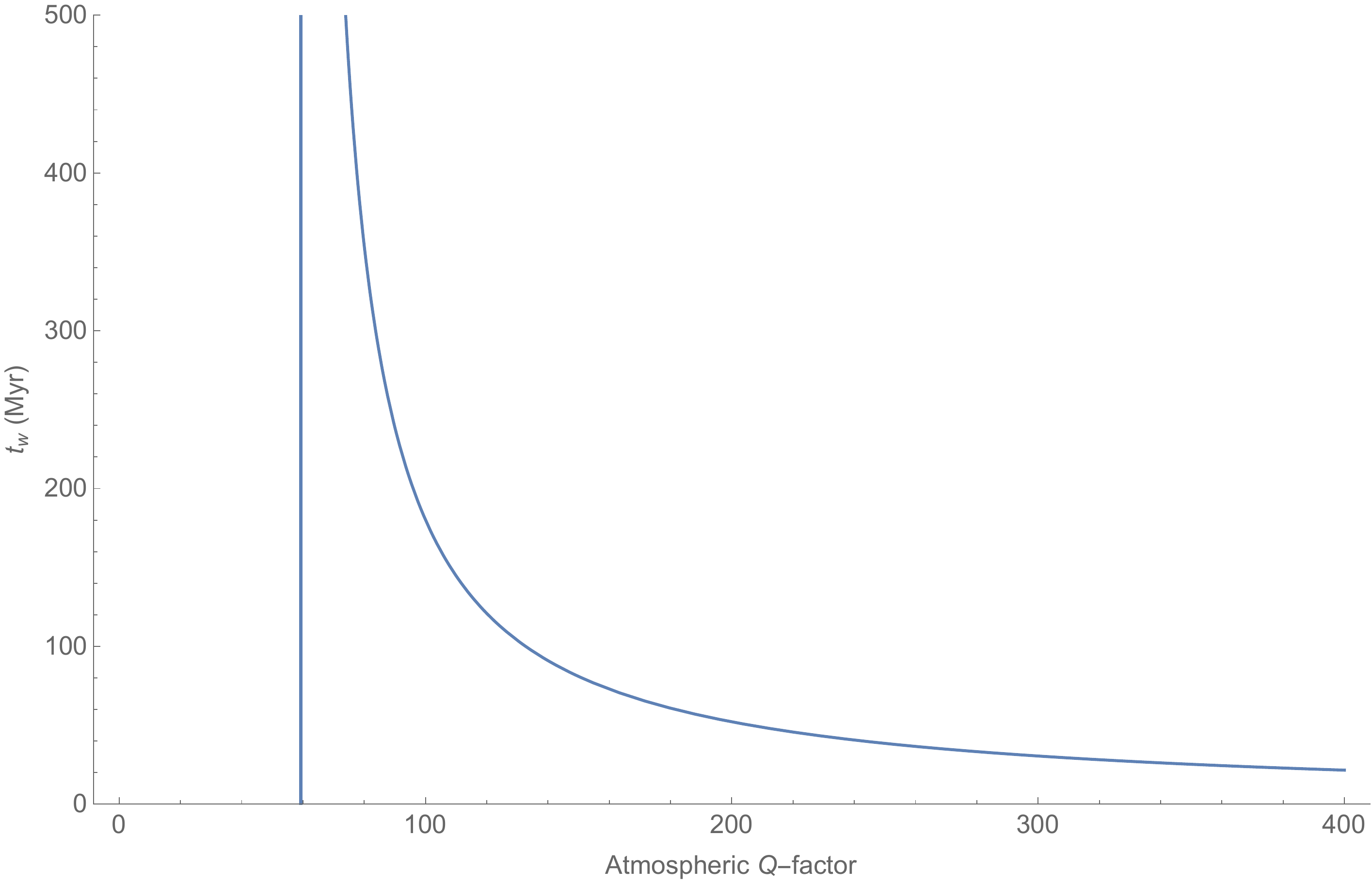}
\caption{The shortest possible stability-preserving warming time $t_w$ as a function of $Q$, as derived in Section 3, with an arbitrary choice of $\Delta T = 10K$.  The asymptote arising at $Q\approx 60$ is a result of no resonance-stabilizing effect occurring, as the maximum value of $\tau_{atm}$ fails to surpass the lunar torque. As this simple model serves only as an upper bound for the conditions required to break resonance, the asymptotic value of $Q\approx 60$ and the timescale $t_w$ will be lower in the computational model, but the general behavior described should still be present.}
\label{figure_label}
\end{figure}

\section{Computational model}

To determine the required conditions for a temperature change to break resonance, a computational model was implemented in Python to numerically compute solutions to the model developed in Section 2 over the $(\Delta T,Q, t_w)$ parameter space.  This generated a stability regime diagram depicting choices of $(\Delta T,Q, t_w)$ resulting in stable and resonance-breaking (unstable) states, shown in Figure 3.

At the program's core is a simulation function which iterates the Earth's rotational frequency as a response to lunar and atmospheric torque as global temperature rises from $T_0 - \Delta T$ to $T_0$ (with $T_0$ being an average global temperature of 287K, though this precise value is unimportant) over an interval of $t_w$ years, simulating the warmup following a period of low global temperatures.

In the absence of a reliable history of the lunar (oceanic) torque, the torque was simulated by taking a "base" value of lunar torque, $\tau_0$, corresponding to the present-day lunar torque (with $\tau_0$ also acting as a scaling factor for Eq. \ref{Aeqn} to convert atmospheric oscillation magnitude to torque) and scaling it by $\tau_{\text{moon}} \propto \tau_0 \cdot \frac{r_0^6}{r^6} \cdot \frac{t}{t_0}$, where $t_0$ is the age of the Earth, $t$ is the progression of the simulation (from $t=0$ to $t=t_0$), and $r_0$ and $r$ are the lunar orbital radii at times $t_0$ and $t$, respectively. The (rather arbitrary) scaling by $\frac{t}{t_0}$ is to loosely approximate the suspected time evolution of oceanic $Q_{oc}$ (and thus $\tau_{\text{moon}}$) over the history of the Earth, as $Q_{oc}$ is thought to have increased over time since the Precambrian. (However, as can be concluded from Figure 5, above some critical lower bound, the actual scaling of the lunar torque has little impact on the evolution of the entire system.)

A very small step size (50yr) was used to minimize numerical error, particularly while simulating at very high $Q$ values.  The simulation function returned whether the result was stable (still trapped in a resonance-stabilizing region) after a warmup period and a subsequent rest period to allow for $\omega$ to settle had passed. To increase computational efficiency, only the stability-instability boundary (the surface shown in Figure 3) was solved for using a multiprocessed binary search, such that the entire simulation ran in a more feasible $\mathcal{O}(n^2 \log n)$ time over the parameter space.

\begin{figure*}[t]
\centering
\noindent\includegraphics[width=\textwidth]{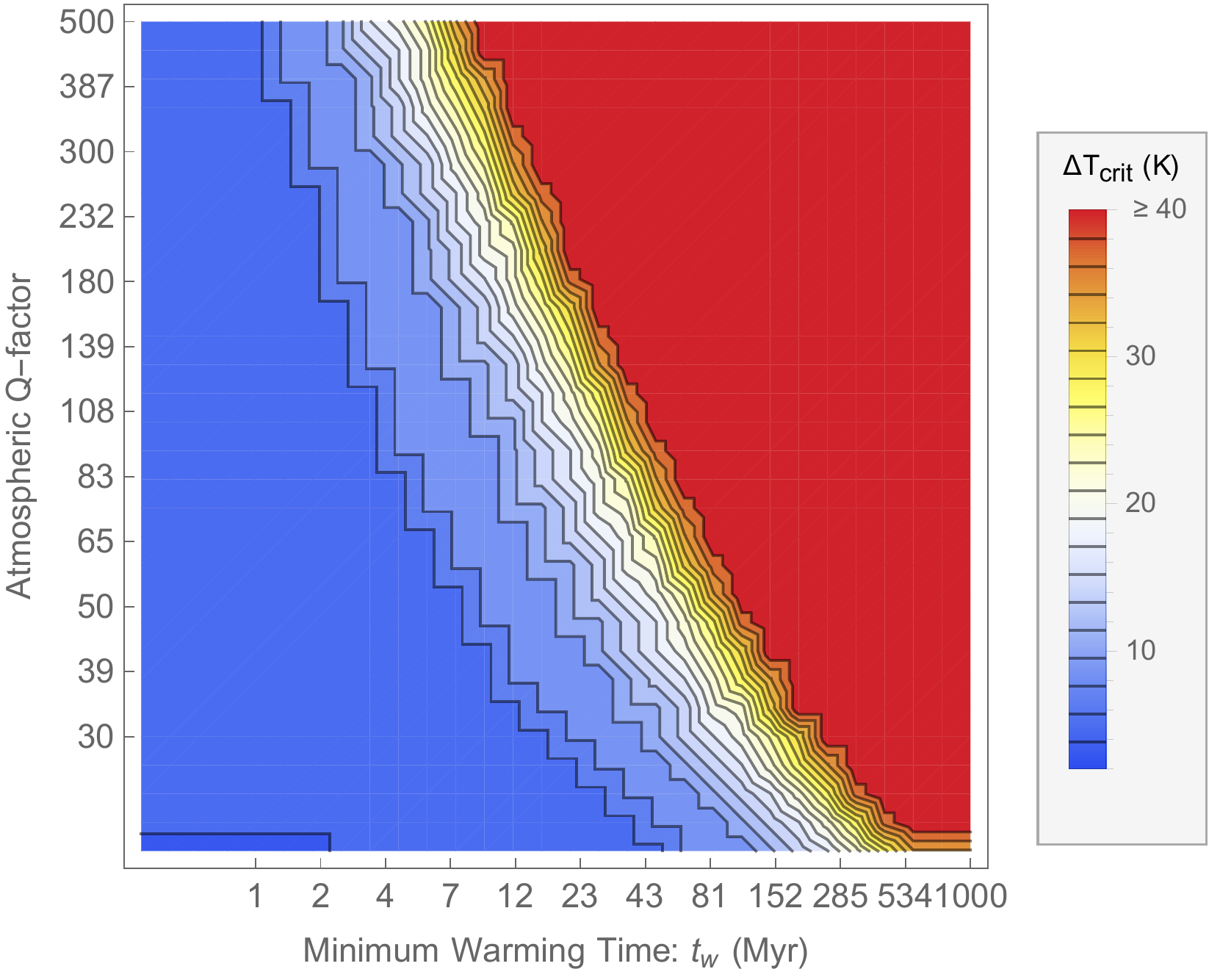}
\caption{The stability-instability boundary calculated along varying $\Delta T$, $Q$, and $t_w$ values. At a fixed $Q$-factor and warming time $(Q_0, t_{w0})$, the resulting critical temperature change $\Delta T_{\text{crit}}$ represents the boundary between stable and unstable resonances: any change $\Delta T_{\text{crit}} + \epsilon$ over the same period of time $t_{w0}$ will break resonance, while any change $\Delta T_{\text{crit}} -\epsilon$ over $t_{w0}$ will preserve resonance. This can be visualized as a bent surface with the $\Delta T$ scale directed out of the page; stable resonances lie below this surface, while unstable resonances lie above the surface.

Higher $Q$-factors permit larger temperature changes per unit time ($\frac{\Delta T}{t_w}$), as the system is more responsive to external torques than scenarios with lower $Q$. Conversely, for a fixed critical temperature change $\Delta T_{\text{crit}}$ over a period of time $t_{w}$, smaller values of $t_w$ require the system to be more responsive to external torques to preserve resonance, requiring a larger $Q$. It should be noted that, regardless of $Q$ and $t_w$, there exists a nonzero minimum value of $\Delta T_0$ required to break resonance (about 5.7 K in the simulation).}
\label{figure_label}
\end{figure*}

\section{Results - $t_w$ timescale}

A regime analysis was performed using our computational model to determine which combinations of atmospheric $Q$, total temperature change $\Delta T$, and warming time $t_w$ resulted in a break of resonance and which preserved resonance, shown in Figure 3.

As expected, for arbitrarily small $t_w$, temperature changes greater than a critical threshold $\Delta T \approx 5 K$ will always break resonance. Above $\Delta T \approx 20$, $t_w$ seems to scale linearly with $\Delta T$. The required $t_w$ to preserve resonance varies inversely with $Q$: with lower $Q$, temperature changes must take place over a larger period of time, as the Earth does not track changes in equilibrium as quickly.  Additionally, the simulation reveals an asymptote for $t_w$ near $Q=10$, with $Q$-factors below this value prohibiting resonance from forming in the first place. All of these behaviors are consistent with the model developed in Section 3.

The overall timescale for the required $t_w$ to break resonance was smaller (by about an order of magnitude) than the rough estimation from Section 3: for a $\Delta T$ of 10K and a $Q$ of 100, temperature changes occurring on a timescale shorter than $10^7$ years would be sufficient to break resonance, while for $Q=30$, as suggested by \citet{lindzen}, changes shorter than 30Myr break resonance, as shown in Figure 3. (It should be noted that while these calculated values of $t_w$ are probably correct to within half an order of magnitude, they are calculated with numerous assumptions and approximations; the relative behavior is more important.) Note that the break in resonance is, of course, conditional on the temperature staying near this increased temperature long enough for the Earth's rotational velocity to decelerate sufficiently away from the area near resonance - a process which would also take on the order of $10^7$ years.  

Thus, our simulations indicate that, ignoring the possibility of recapture, had the rotational velocity and temperature of the Earth previously reached an equilibrium during a "snowball Earth", virtually any realistic subsequent deglaciation period would break resonance, as discussed further in Section 8.

\section{Results - effects of thermal noise on resonant stability}
In addition to a systematic global climate change following a cool-constant-warm pattern like a snowball event, the computational model outlined in Section 4 was also further developed to test the resilience of the resonance to random atmospheric thermal noise: higher-frequency fluctuations occurring at a variety of amplitudes. The temperature was driven sinusoidally across a very large range of frequencies and amplitudes encompassing all reasonable values for small-scale global temperature fluctuations.  These results are detailed in Figure 4.  It was found that, for a sinusoidally driven global atmospheric temperature, the optimal fluctuation period to break resonance - that is, the frequency whereby the required amplitude to break resonance is minimized - was on the order of 10000 years.  However, the required thermal amplitude for this value was approximately 20K (half-wave amplitude, so a total temperature oscillation of 40K), which is unrealistic, so the possibility of resonant break due to random thermal fluctuations was discarded. Further evidence for discarding this possibility is also provided by the results from the final figure in this paper.

\begin{figure}[b]
\centering
\noindent\includegraphics[width=.9\columnwidth]{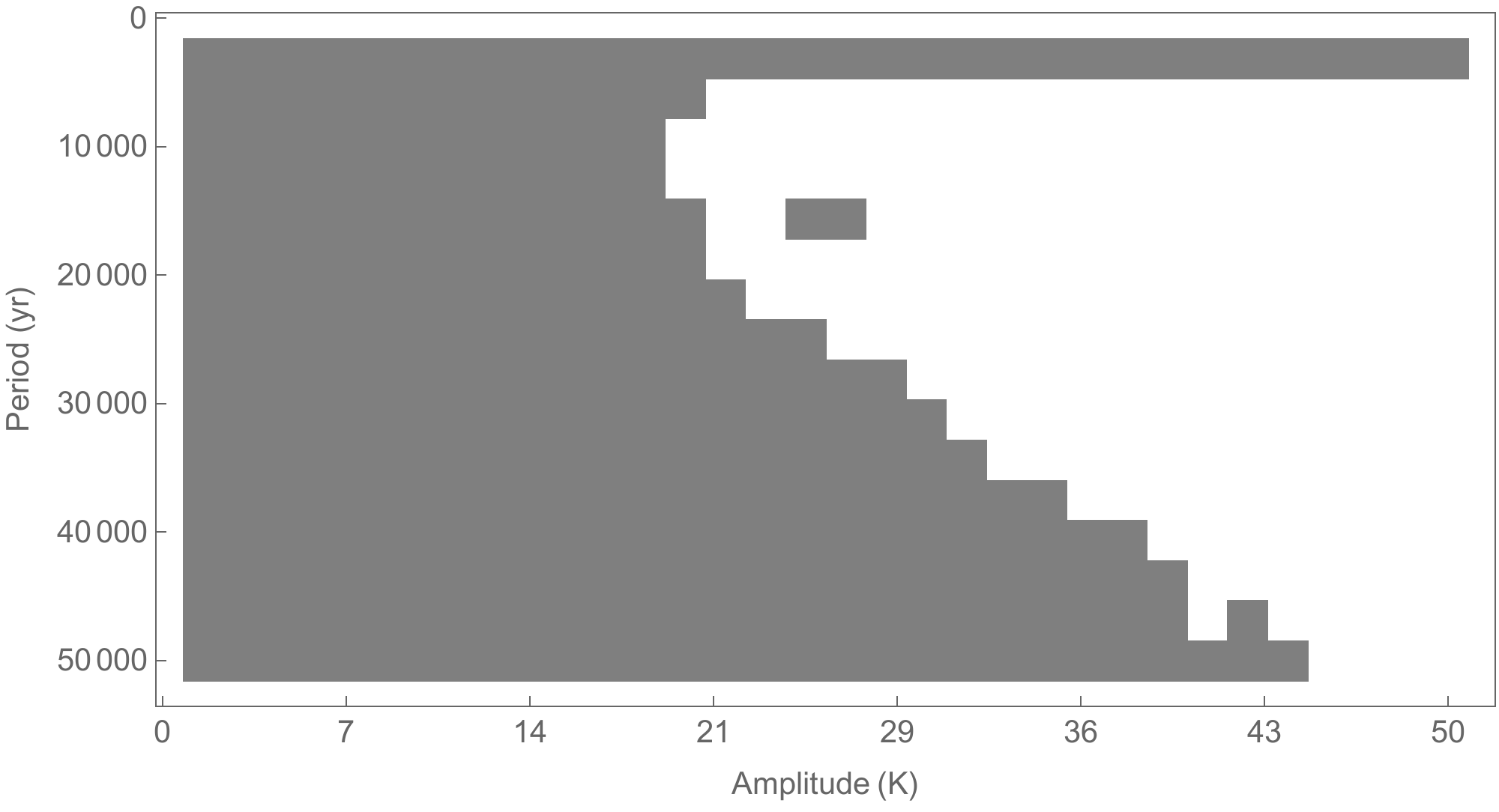}
\caption{Regime analysis of sinusoidally driven atmospheric temperature fluctuations across (half-wave) amplitude and frequency for an initial phase of zero.  Grey regions indicate resonance-preserving scenarios, while white regions break resonance. The "noise" in the diagram, such as the small island of stability in the white region is due to the fact that breakage also depends weakly on initial phase of the sinusoidal driver. However, phase was found not to change the overall shape of the curve, aside from small changes near the edge, so the resilience of the atmospheric resonance to realistic thermal noise is independent of phase.}
\end{figure}

\section{Results - simulated length of day over time}

Finally, we used the model from the above two sections to create a simulation of Earth's length of day over its history, shown in Figure 5.  Given the plausibility of a snowball event breaking resonance, we simulated a sequence of four snowball events, corresponding in time and duration to  an early Paleoproterozoic glaciation discussed in \citet{kirschvink} and three possible "snowball Earth" events during the late Precambrian: the Kaigas, the Marinoan, and the Sturtian glaciations. It should be mentioned that the time and duration (and recently, even the validity \citep{rooney2}) of the Kaigas glaciation are debated; it is included at ~920Ma and given a short duration to simulate resonant recapture for demonstrative purposes. A variety of base torque values ($\tau_0$)were chosen for simulation. Throughout the simulation, random atmospheric noise was also simulated as the sum of several sinusoidal drivers with a maximum amplitude of approximately 5K. (This temperature curve is arbitrary and should not be interpreted as an actual thermal history of the Earth's atmosphere, which is not fully agreed upon; it was generated to demonstrate the main points of this study.)

Existing stromatolite data as compiled in \citet{williams} put the point of resonant breakage near 600Ma., while the data points at approximately 2Ga., though even less reliable, could very tentatively establish a lower bound on the formation of this resonance. After 600Ma, stromatolite, coral, and bivalve data indicate that the day length increases to its current 24 hours day length quickly after a period of relatively constant day length (though paleorotational data is nearly absent during most of this range, only available near the endpoints). However, this data, particularly the early stromatolite data \citep{panella}, should not be taken too seriously. \citep{zahnle} Paleontologists \citet{scrutton}, and \citet{hoffmanStromatolites} also found this data to be unreliable and unsuitable for precise quantitative analysis. Regrettably, no significant additional data has emerged in the past several decades; further and more reliable data will be needed to test both Zahnle's and Walker's hypothesis and our developments on mechanisms of breaking resonance.

\section{Conclusions}

Our model supports the first scenario presented in the introduction: the Earth entered a resonant state, perhaps at about 2Ga. before present (though this value is highly uncertain, as it depends on an unknown evolution of lunar oceanic $Q_{oc}$ for that epoch which is only crudely simulated in our model). The Earth then escaped resonance at about 600Ma. (this value also depends on $Q_{oc}$), when resonance was broken by a global temperature increase that is well explained as the deglaciation following a snowball event.

As shown in the Figures 2 and 3, an asymptote dependent on lunar torque exists such that there is a critical value of atmospheric $Q$ below which resonance will not form. Near this value, the resonance is quite unstable. Computationally, this asymptotic value was found to be very low: $Q\approx 10$ for the present lunar torque, and possibly lower for some smaller estimated Precambrian lunar torques \citep{zahnle}, making resonance formation likely given an estimate of $Q\approx 30$ from \citet{lindzen}.

The minimum warming time $t_w$ required to break a resonance state was found to be within values that would be broken by a deglaciation following a snowball event; for most values of $Q$, the deglaciation period would need to be less than $10^7$yr., easily within the $t_w$ estimates for snowball deglaciations presented by \cite{hoffman}. Snowball events with depressed, relatively stable temperatures lasting for a period of around $10^7$ years (also similar timespans as in \cite{hoffman}) were found to provide sufficient time for an equilibrium of $\omega$ and $\omega_0$ to be reached such that the subsequent deglaciation breaks resonance, though this value also depends inversely with the lunar torque, which is not accurately simulated over time in our model.

The mid-Precambrian was lacking in global or near-global glaciations, with the possible exception of the Huronian glaciation ca. 2.4-2.2 Ga.\citep{melezhik}, which likely occurred before resonance had formed. Similar early Paleoproterozoic glaciation was argued to be a "snowball event" in \citet{kirschvink}; however, the parameters described in that study were simulated and resulted in resonant recapture. The fact that there is little evidence of any potentially resonance-breaking glaciation for almost a billion years prior to the Sturtian glaciation \citep{rooney} lends credence to the idea that the deglaciation of a "snowball Earth" was the likely trigger that broke resonance after allowing it to persist for a length of time on the order of a billion years.

It should be noted that while a reasonable choice of atmospheric and lunar variables makes the scenario described in this study possible and likely, the paleorotational data available is not sufficient to confirm the hypotheses of resonance formation or breakage. Further data is required; it is our hope that this work will encourage developments in this area. 

\begin{figure*}
\centering
\noindent\includegraphics[width=\textwidth]{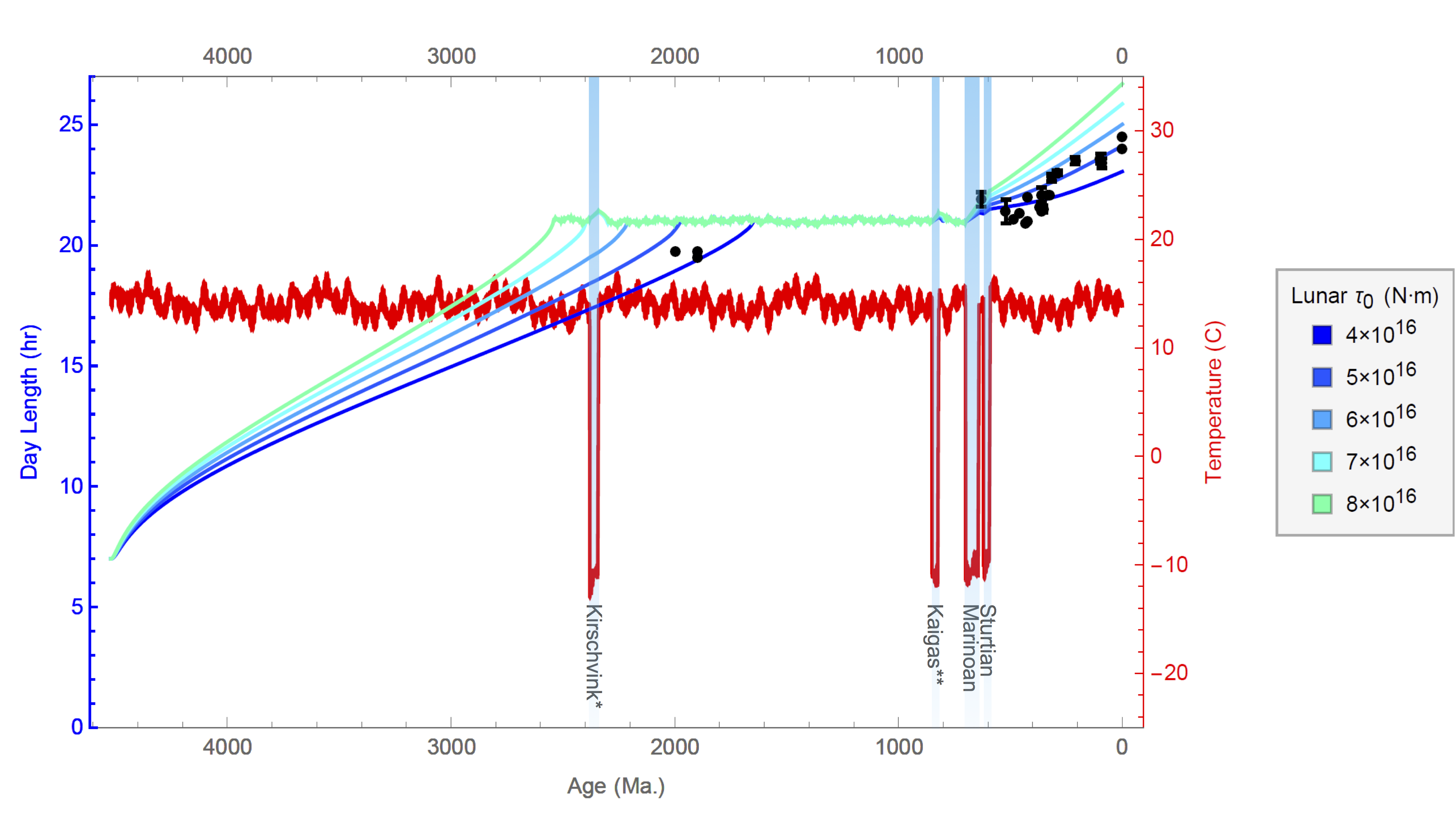}
\caption{Day lengths with varying choices of $\tau_0$ (shown in blues and greens) and temperature values (red) over the lifetime of the Earth. Note that atmospheric thermal noise does not influence the day length except very near resonance, and that the resonant effect remains unbroken through this noise until two successive simulated snowball events at the end of the Precambrian 720Ma. and 640Ma., corresponding to recent estimates of the Sturtian and Marinoan glaciations. \citep{rooney}  Recapture events can be seen at 870Ma following the "Kaigas glaciation" and, for some values of $\tau_0$, following a Paleoproterozoic glaciation detailed in \citet{kirschvink}. Approximate empirical day length data from a compilation in \citet{williams} are overlayed in black (error bars included where present), and resemble the simulated results, though the reader should not take this data to be too reliable, particularly the data points prior to 600Ma.}
\label{fig5}
\end{figure*}

\appendix

\section{Source code}

All of the code used in this paper is available upon request from the first author.


%
%
%
%
%
%
%

\begin{acknowledgments}
The authors wish to thank the late Tom Tombrello for several contributions to this paper. We also thank two reviewers for their constructive feedback.
\end{acknowledgments}

\end{article}
%
%
%
%
%
%
%
%



\begin{thebibliography}{}

\bibitem[Hide, et al. (1996)]{hide} Hide, R., Boggs, D.H., Dickey, J.O., Dong, D., Gross, R.S. and Jackson, A., 1996. Topographic core-mantle coupling and polar motion on decadal time-scales. Geophys. J. Int., 125, 599–607.
\bibitem[Hofmann (1973)]{hoffmanStromatolites} Hofmann, H.J., 1973. Stromatolites: Characteristics and utility, Earth-Sci. Rev., 9, 339-373. 
\bibitem[Hofmann and Schrag (2002)]{hoffman} Hofmann, P., and Schrag, D., 2002. The "snowball Earth" hypothesis: testing the limits of global change. Terra Nova, 14: 129-155.
\bibitem[Kirschvink, et al. (2000)]{kirschvink} Kirschvink, J.L. et al., 2000. Paleoproterozoic snowball Earth: Extreme climatic and geochemical global change and its biological consequences. PNAS, 97 (4), 1400-1405. 
\bibitem[Lamb (1932)]{lamb} Lamb, H., 1932. Hydrodynamics. Dover, New York, 738pp.
\bibitem[Lambeck (1980)]{lambeck} Lambeck, K., 1980. The Earth's Variable Rotation. Cambridge University Press, Cambridge, 500pp.
\bibitem[Lindzen and Blake (1972)]{lindzen} Lindzen, R.S., and Blake, D., 1972. Lamb waves in the presence of realistic distributions of temperature and dissipation. J. Geophys. Res., 77(12), 2166–2176.
\bibitem[Lindzen and Chapman (1969)]{lindzenandchapman} Lindzen, R.S., and Chapman, S., 1969. Atmospheric tides. Sp. Sci Revs., 10, 3-188.
\bibitem[Melezhik (2006)]{melezhik} Melezhik, V.A., 2006. Multiple causes of Earth's earliest global glaciation. Terra Nova, 18:130-137.
\bibitem[Panella (1972)]{panella} Pannella, G., 1972. Precambrian stromatolites as paleontological clocks. Internat. Geol. Congr. 24th Session, Montreal, Proc. Section 1, 50-57. 
\bibitem[Pierrehumbert, et al. (2011)]{pierrehumbert} Pierrehumbert, R.T., Abbot, D.S., Voigt, A., and Koll, D., 2011. Climate of the Neoproterozoic. Annu. Rev. Earth Planet. Sci. 2011. 39:417–60
\bibitem[Rooney, et al. (2014)]{rooney} Rooney, A.D., Macdonald, F.A., Strauss, J.V., Dudás, F.Ö., Hallman, C., and Selbye, D., 2014. Re-Os geochronology and coupled Os-Sr isotope constraints on the Sturtian snowball Earth. PNAS, 111(1), pp.51-56.
\bibitem[Rooney, et al. (2015)]{rooney2} Rooney, A. D., J. V. Strauss, A. D. Brandon, and F. A. Macdonald, 2015. A Cryogenian chronology: Two long-lasting synchronous
Neoproterozoic glaciations. Geology 43:459–462.
\bibitem[Scrutton (1978)]{scrutton} Scrutton, C.T., 1978. Periodic growth features in fossil organisms and the length of the day and month. Tidal Friction and the Earth's Rotation, Brosche, P. and Sündermann, J. (eds), Springer, Berlin, 154-196
\bibitem[Williams (2000)]{williams} Williams, G., 2000. Geological constraints on the Precambrian history of Earth's rotation and the Moon's orbit. Rev. Geophys., 38:37-59.
\bibitem[Zahnle and Walker (1987)]{zahnle} Zahnle, K. and Walker, J.C.G., 1987. A constant daylength during the Precambrian Era? Precambrian Res., 37:95-105.


%
%
%
%

\end{thebibliography}
\end{document}